\documentclass[twocolumn,showpacs,preprintnumbers,amsmath,amssymb]{revtex4-1}

\usepackage{graphicx}
\usepackage{dcolumn}
\usepackage{bm}
\usepackage{float}
\usepackage{float}
\begin{document}

\preprint{APS/123-QED}

\title{Memory-Enhanced Noiseless Cross Phase Modulation}
\author{M. Hosseini$^1$, S. Rebi\'c$^{1,2}$, B. M. Sparkes$^1$, J. Twamley$^2$, B. C. Buchler$^1$, P. K. Lam$^1$}

\affiliation{$^1$ Centre for Quantum Computation and Communication Technology, Department of Quantum Science, The Australian National University, Canberra, Australia \\
$^2$ Centre for Engineered Quantum Systems, Department Physics $\&$ Astronomy, Macquarie University, North Ryde, NSW 2109, Australia}

\date{\today}

\begin{abstract} 

Using a gradient echo memory, we experimentally demonstrate cross phase modulation (XPM) between two optical pulses; one stored and one freely propagating through the memory medium. We explain how this idea can be extended to enable substantial nonlinear interaction between two single photons that are both stored in the memory. We present semi-classical and quantum simulations along with a proposed experimental scheme to demonstrate the feasibility of achieving large XPM at single photon level.
\end{abstract}

\maketitle

 The optical Kerr effect is present in most materials but only becomes significant with very intense optical fields and/or long interaction times. In the limit of extreme nonlinearity, individual photons could be persuaded to interact strongly with one another and induce cross-phase modulation (XPM). This kind of interaction is the basis of the control-not gate and phase-not gate that lie at the heart of quantum computing algorithms~\cite{Milburn:PRL1989,Sebastien:NJP:2007}. In addition, a large XPM can also be used for generation of cluster state as building block of one-way quantum computing~\cite{Briegel:PRL:2001,*Raussendorf:PRL:2001,*Walther:nat:2005}, as well as nonlinear optical switching~\cite{Williams:nlsw:1984}.
  

 To date, there have been various proposals aimed at bringing this strong interaction to existence. Optical fibers are an attractive XPM medium~\cite{Kerrinfibre:nph:2010}. While they may not be highly nonlinear, the interaction times can be extended simply by using longer fibers. Unfortunately, the fast response time of optical fibers makes it impossible to temporally mode match two co-propagating single photons~\cite{Shapiro:PRA:2006}. Even if the nonlinearity is large, the near-instantaneous response of the medium implies that only a small and randomly distributed portion of the single photon wave packet experiences the phase shift~\cite{Shapiro:NJP:2007}. 
 
Another method of facilitating long interaction time for XPM is via light interaction with an atomic ensemble. Electromagnetically induced transparency (EIT) based XPM exploits slow-light effects in an atomic ensemble to enhance the nonlinear interaction of light fields via the ac-Stark effect \cite{Kerr-EIT}. A phase shift almost two orders of magnitude larger than that in optical fibers has been observed using EIT~\cite{Hsiang:PRL:XPM:2011,*Shiau:PRL:2011}. It has, however, been theoretically demonstrated that EIT-based XPM suffers from severe loss in regimes where large phase shifts are expected~\cite{Banacloche:PRA:2010}. Moreover, self phase modulation (SPM) poses a fundamental limit to the fidelity of the output states~\cite{Jonecki:SPM:1993}.

 Memory-based XPM offers a solution to the temporal mode-matching issue. One pulse can be stored in the atomic coherence while a second pulse  propagates through the memory medium allowing for XPM~\cite{Chen:ExpEIT:XPM:PRL:2006}. In this paper, we first report a proof of principle experiment that shows existence of XPM inside a Raman gradient echo memory ($\Lambda$-GEM)~\cite{HetetPRL:100:2008}. We then propose a more elaborate scheme that can potentially allow large XPM between single photons simultaneously stored in the memory. 

 In the $\Lambda$-GEM scheme, a weak optical field is coherently Raman coupled into the spin coherence of an ensemble of three-level atoms using a strong coupling field~\cite{HetetPRL:100:2008}. A linearly varying ground-state splitting is applied using a magnetic field gradient, $\eta$, such that the light pulse is decomposed into its constituent Fourier frequencies that are then stored longitudinally along the storage medium. Reversing the sign of the gradient will time-reverse the absorption leading to a coherent photon echo. This storage scheme has been proven to be versatile, efficient and noiseless~\cite{Hosseini:2009p8466, *Hosseini:natcomm:10, *Hosseini:NatPhys:2011}.

To understand the dynamics in GEM, we use a polariton that is a superposition of the electric field, $\hat{\mathcal{E}}$, and atomic spin coherence, $\hat{\sigma}_{12}$, in the spatial Fourier domain~\cite{PRL-3GEM}, defined as $\hat{\psi}(t,k)=k\hat{\mathcal{E}}(t,k) + \mathcal{N}\Omega_c/\Delta\hat{\sigma}_{12}(t,k)$, where $k$ is the spatial frequency, $\mathcal{N}$ is the effective linear atomic density, $\Delta$ is the Raman detuning from the excited state, and $\Omega_c$ is the coupling field Rabi frequency. During storage, the polariton evolves to higher $k$-values at a rate proportional to $\eta$. When  using GEM for XPM, it is the polariton that will be phase shifted, leading to a phase shift of the photon echo on recall from the memory.

There are three properties of the polariton that are important to the following discussion: i) The Fourier transform of the Maxwell equation gives $k\hat{\mathcal{E}}(t,k)=\mathcal{N}\hat{\sigma_{12}}(t,k)\Omega_c/\Delta$ \cite{PRL-3GEM}. Because the spin coherence has a constant amplitude during storage, the Maxwell equation implies that $\mathcal{E}$ is inversely proportional to $k$. ii) The polariton can be stopped in $k$-space by switching $\eta$ to 0. For a pulse stored with $\eta=0$, the group velocity of the optical field is found to be $v_g=g\mathcal{N}/k^2(\Omega_c/\Delta)^2$. iii) The polariton is purely atomic, $|\hat{\mathcal{E}}|=0$, when the coupling field is off.

\begin{figure}[!t]
\centerline{\includegraphics[width=\columnwidth]{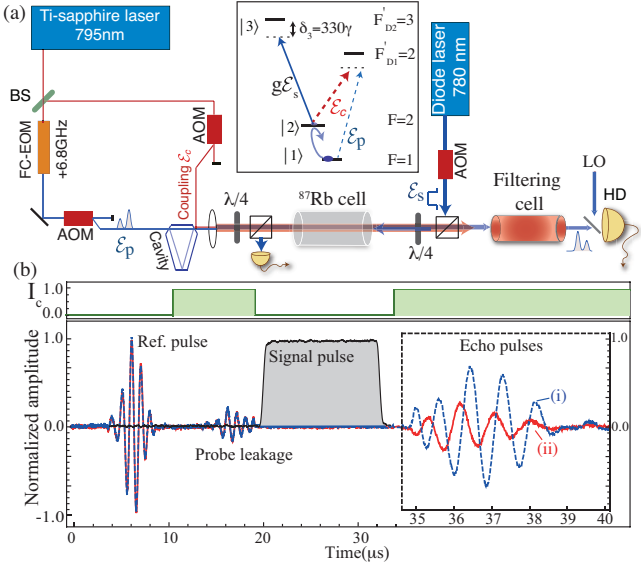}} 
 \caption{(a) Schematic experimental setup. The probe field ($\mathcal{E}_p$) is shifted by 6.8~GHz with respect to the coupling field ($\mathcal{E}_c$) using a fiber-coupled electro-optic modulator (FC-EOM) and is combined with the coupling field using a ring cavity. The signal field ($\mathcal{E}_s$) is counter-propagating with other beams. The inset shows the atom-light interaction scheme. (b) Heterodyne data showing normalized amplitude of the modulated phase reference, input and echo probe pulses. The top trace shows the switching protocol of the coupling field intensity. Traces (i) (blue) and (ii) (red) show the amplitude of $\mathcal{E}_p$ measured at the output of the memory without and with the signal pulse, respectively. Trace (ii) is taken 60~$\mu$s after (i) and overlapped using the reference pulse as a timing signal.}
  \label{setup}
 \end{figure}

 The experimental setup is shown in Fig.~\ref{setup} (a). A coupling field and a weak probe field were combined using a ring resonator and sent into a 20 cm long gas cell of $^{87}$Rb mixed with 0.5 Torr of Kr buffer gas. The gas cell was housed inside a pair of solenoids that generate switchable magnetic field gradients. After the memory, the coupling field was suppressed using a  $^{85}$Rb gas cell and the probe beam was measured using heterodyne detection (see Ref.~\cite{Hosseini:NatPhys:2011} for more experimental details).
 
The probe was stored for approximately 15~$\mu$s while the coupling field was switched off. During that time, a signal field generated from a diode laser and detuned by $\delta_3\simeq2$~GHz from $F=2\to F^{\prime}=3$ of $^{87}$Rb $D_2$ line was sent through the memory. This field was counter-propagating with respect to the probe and coupling fields to avoid measurement contamination. The signal field gives rise to an ac-Stark shift of the spin coherence. On recall, therefore, the stored probe field will be phase shifted proportional to the strength and duration of the signal field. To measure the size of the phase shift, we ran two storage experiments in quick succession, without and with the signal field, as shown in Fig.~\ref{setup} (b) (i) and (ii) respectively. A phase reference for the two recalled probe pulses was provided by a pulse that passed through the memory cell 10~$\mu$s before the start of each experiment. This reference pulse then allowed us to compare the recalled probe phase with and without the signal field, as indicated in the figure.
  \begin{figure}[!t]
\centerline{\includegraphics[width=\columnwidth]{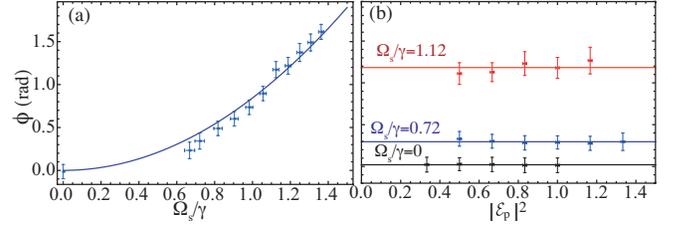}} 
 \caption[XPM experiment]{(a) Nonlinear phase shift as a function of signal pulse Rabi frequency. The solid line is the predicted theory curve. (b) Phase shift for different signal pulse Rabi frequency as a function of peak intensity of the probe pulse normalized to the reference pulse.}
  \label{exp}
 \end{figure}  
 
To verify that this phase shift is due to a nonlinearity in the memory, we measured the phase shift as a function of the signal field Rabi frequency, as shown in Fig.~\ref{exp}(a), where the solid line represents the theoretical expectation calculated~\cite{Chen:ExpEIT:XPM:PRL:2006} using $\phi_\mathrm{XPM}=\Omega_s^2\delta_3\tau/2(\gamma^2+\delta_3^2)$, where $\Omega_s=g|\hat{\mathcal{E}}_s|$ is the signal field Rabi frequency and $\tau$ is its duration. Crucially, our scheme has no measurable SPM, as shown in Fig.~\ref{exp} (b), where the recalled probe phase is seen to be independent of the probe intensity.
 
It is interesting to compare our scheme to EIT, since the atomic medium and experimental setup are very similar. It has been shown theoretically that EIT based XPM suffers from self-phase modulation and severe loss in the regimes where large phase shifts are expected~\cite{Banacloche:PRA:2010}. This is because some frequency components of one light field, lying outside the EIT window interact with all the components of the other light field thus driving the system out of the dark state. In our scheme, the bandwidth of the GEM can be tuned to match the bandwidth of the probe field without otherwise changing the inherent nonlinearity of the system.
 
Based on the experimental data presented in Fig.~\ref{exp} (a), we estimate a phase shift on the order of $10^{-12}$~rad for signal and probe fields containing single photons. In our experiment,  the large detuning of the signal field (2~GHz) severely reduces the available nonlinearity, but it is necessary due to the large Doppler broadening of thermal atoms. In fact, even with this detuning the scattering due to the signal field leads to substantial loss of the atomic coherence. In Fig.~\ref{setup}(b), for example, the probe recall is reduced from 53\% to 7\% by the signal field. In cold atomic ensembles~\cite{Chen:ExpEIT:XPM:PRL:2006} this detuning could be reduced by two orders of magnitude allowing two orders of magnitude larger phase shift. Even if larger phase shifts can, in principle, be achieved using cold atomic samples, this particular nonlinear interaction scheme might not be useful for single photon interactions. The signal field is not stored in the memory, meaning that the interaction time with the probe will be limited. We will now analyze a scheme in which the probe and signal fields are simultaneously stored in a double-GEM system. As before, the origin of this XPM is the ac-Stark effect, but the the available phase shift can be increased by increasing the interaction time.
\begin{figure}[!t]
\centerline{\includegraphics[width=0.95\columnwidth]{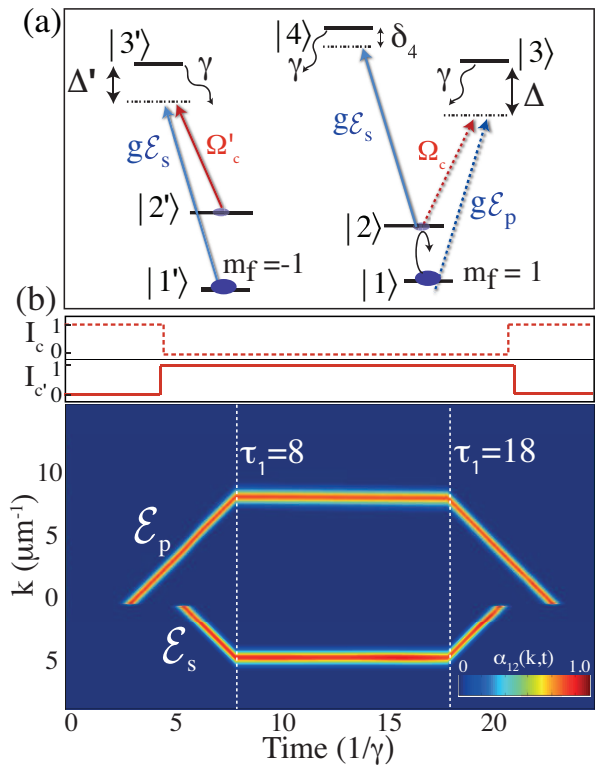}}
 \caption[Level scheme for XPM in memory]{ (a) Schematic atomic level structure of $^{87}$Rb showing a scheme for the proposed nonlinear interaction. The probe $\mathcal{E}_p$ and signal $\mathcal{E}_s$ pulses arrived at different times are independently mapped to atomic coherence $|1\rangle\langle 2|$ and $|1'\rangle\langle 2'|$ using two coupling fields $\Omega_{c}$ and $\Omega'_{c}$,  respectively. The signal field can modifies the phase of the atomic coherence $|1\rangle\langle 2|$ via the ac-Stark effect. (b) The total atomic coherence in spatial Fourier space ($k$) and time representing the evolution of two atomic fields in $k-t$ plane. The gradient field is switched off during $\tau_1=8<t<\tau_2=18$. The top part of the figure shows the coupling field switching protocols.}
   \label{levelst}
 \end{figure}  

The Rb level structure and nonlinear interaction scheme between two photons stored inside the memory are shown in Fig.~\ref{levelst} (a). The two fields (with Rabi frequencies of $g\hat{\mathcal{E}}_p$ and $g \hat{\mathcal{E}}_s$) are stored independently in two atomic spin coherences, $\hat{\sigma}_{12}$ and $\hat{\sigma}_{1'2'}$, using two coupling fields with Rabi frequencies of $\Omega_c$ and $\Omega'_{c}$. 

The scheme works as shown in Fig.~\ref{levelst} (b). The probe and signal enter the medium consecutively. By timing the two coupling fields, each pulse is independently mapped into distinct polaritons. These modes propagate in opposite directions in $k$-space due to the opposite sign of the $m_f$ in the hyperfine states. The evolution of the two polaritons is stopped at a constant $k$ by switching $\eta$ to zero soon after the signal pulse enters the medium. The coupling field for the probe is then switched off mapping the polariton into the spin coherence $\hat{\sigma}_{12}$. The coupling field for the signal is left on, ensuring a photonic component of this polariton that will give rise to an ac-Stark shift of the probe. The maximum available interaction time is proportional to the medium length and inversely to the group velocity of the signal light. After a controllable interaction time, the frequency gradient and coupling field $\Omega_c$ can be switched to recall the probe, which will be phase shifted due to the interaction with the signal field.
We now analyze the system with both semi-classical and quantum simulations. The interaction Hamiltonian of the system with level scheme depicted in Fig.~\ref{levelst} (a) can be written as
\begin{eqnarray}
H_I=\frac{N}{L}\int[g\hat{\mathcal{E}}_p\hat{\sigma}_{31}+\Omega_{c}\hat{\sigma}_{32}+g\hat{\mathcal{E}}_s\hat{\sigma}_{42} \nonumber\\
+g\hat{\mathcal{E}}_s\hat{\sigma}_{3'1'}+\Omega'_c\hat{\sigma}_{3'2'}+H.c]dz
  \label{eq:Hamilt}
 \end{eqnarray}
 where $\hat{\sigma}_{ij}$ are the collective atomic spin operators, $g$ is the atom-field coupling constant, $N$ is total number of atoms in the interaction volume and $L$ is the ensemble length. We numerically solve the Maxwell-Bloch equations in semiclassical regime by replacing operators by $c$-numbers. The phase of the coherence $\hat{\sigma}_{12}$ after the interaction time is then  given by
\begin{eqnarray}
\phi_\mathrm{XPM}=\int_{\tau_1}^{\tau_2}{\frac{|g\hat{\mathcal{E}}_s(t-\tau_1)|^2 \delta_4}{\gamma^2+\delta_4^2}dt}
 \end{eqnarray}
and the loss rates of $\hat{\sigma}_{1'2'}$ and $\hat{\sigma}_{12}$ are given by $\gamma(\Omega'_c/\Delta')^2$ and $|g\mathcal{E}_s(t-\tau_1)|^2 \gamma/(\gamma^2+\delta_4^2)$, respectively. Our semi-classical numerical simulation predicts a phase shift of about 0.02~mrad per photon after interaction time of $15/\gamma$. We observed that the XPM is invariant with the probe amplitude, again showing immunity to SPM. 
 To characterize the system in quantum regime we perform quantum simulations by solving the master equation numerically. It is assumed that initially a photon is encoded in a coherence between $|1\rangle$ and $|2\rangle$ so the initial state of the atomic system becomes $\rho_{at} = 1/2 \left( |1'\rangle\langle 1'| + |\psi_0\rangle\langle \psi_0|\right)$, where $|\psi_0\rangle = (|1\rangle+|2\rangle)/\sqrt{2}$. The initial state of the incoming signal photons is then given by $\rho_{ph} = |0_p,\psi_s\rangle\langle 0_p,\psi_s|$, where $|\psi_s\rangle = \left( |0_s\rangle + |1_s\rangle\right)/\sqrt{2}$, giving the total initial state $\rho(0) = \rho_{at} \otimes  \rho_{ph}$.
 
\begin{figure}[!t]
\centerline{\includegraphics[width=\columnwidth]{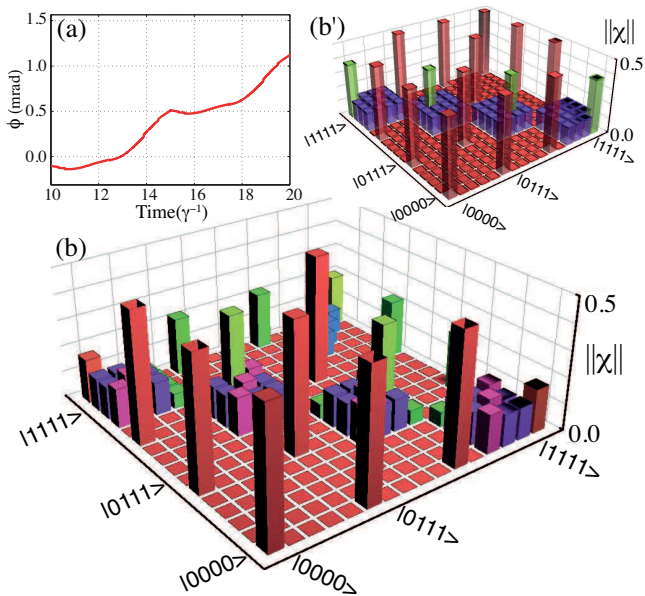}}
\caption{(a) Conditional Phase shift $\phi$. The following parameters were used $\Omega_{c}=\Omega'_c = 20\gamma$, photon bandwidth$=\gamma$, $\Delta =\Delta'= 30\Omega_{c}$, $\delta_4 = 20\gamma$, $g = 0.085\gamma $. The light is coupled to the atomic medium with $g_{13} = g\sqrt{N}$, $g_{24} = g$ and $g_{1'3'} = g\sqrt{N}$, where $g_{ij}$ is the effective coupling strength between $|i\rangle\to|j\rangle$, $N = 10^7$ is the number of atoms. For time $t<10/\gamma$ the system is in the transient regime. (b) Choi matrix representing quantum tomography of the gate operation inside the memory. The Choi matrix of an ideal gate operation is also shown in (b$'$).}
\label{CPS}
\end{figure}

Next, we solve the master equation including terms describing decoherence due to the spontaneous decay from the excited states~\cite{Walls:Milburn:QO}. From the resulting density operator, the conditional phase shift between a single photonic qubit in mode $s$ and polaritonic qubit encoded in atomic coherence $|\psi_0\rangle$ is calculated as a function of interaction time. Results for the conditional phase shift $\phi$ and the accompanying gate fidelity are shown in Fig.~\ref{CPS} (a) for a set of parameters closely corresponding to the experiment. We note that the predicted phase shift here is of the same order of magnitude as in the semiclassical simulation. To obtain a more general characterization of the gate process, we analyze a point where the time of interaction is $15/\gamma$, and perform quantum process tomography~\cite{Benenti09} which, in effect, averages over all possible input states of the two qubit system. The result, in the form of a Choi matrix ($\chi$), is shown in Fig.~\ref{CPS}(b). The ideal gate operation Choi matrix is shown in Fig.~\ref{CPS}~(b$'$). The process fidelity, defined as trace overlap ${\mathcal F} = \textrm{Tr}\{\chi_{ideal}\cdot \chi\}$ is calculated to be $\sim$85\%. 
In the proposed scheme, the far-detuned interactions as well as tuneability and control over the strength and duration of the interaction could potentially resolve the temporal mode-matching issue associated with other proposed XPM schemes such as EIT. Although the nonlinear phase shift for the proposed scheme is smaller than $\pi$, the current scheme can be used to implement parity or phase gates where the strength of the coherent states can offset the weakness of the nonlinearities~\cite{Munro:NJP:2005, Sebastien:NJP:2007}.\\
There are also routes to further enhancement of the nonlinearity. In particular, the interaction strength is limited by the storage of the signal field which still has a non-zero group velocity in the memory. The application of a counter-propagating coupling field for the signal field would allow stationary trapping of the signal light. The physics of the trapping light in this case is similar to stationary light generated in an EIT medium that has previously been experimentally demonstrated~\cite{Lukin-EIT-Nature}. The stationary light in GEM is generated through an off-resonance interaction and therefore suffers less loss compared to the EIT case. The study of stationary light in GEM is beyond the scope of this work and will appear in future publications. Our simulations also suggest that XPM on the order of $\pi$ at single photon level might only be feasible in cavity BEC systems~\cite{Colombe:CBEC:Nature:2007} due to their small atom-light interaction volume, large coupling strength and long coherence times. 
 
We conclude by noting that, in addition to the demonstrated efficient quantum storage and capability to arbitrarily manipulate optical pulses, the versatility of GEM can be extended to implement a parity gate from which CNOT gate can be constructed \cite{Munro:NJP:2005}. The lack of SPM and the demonstrated noiseless high efficiency storage in our scheme suggests that the proposed method could be a potential candidate for implementing practical XPM between single photons and it can be applied to various concepts of quantum information.

We thank John Close, Andrew White and Andr\'e Carvalho for enlightening discussions. This research was conducted by the Australian Research Council Centre of Excellence for Quantum Computation and Communication Technology (project number CE110001027) and Centre of Excellence for Engineered Quantum Systems (Project number CE110001013).

 \bibliographystyle{plainbib}
\bibliography{ref6}
 \end{document}